\documentclass[journal=jacsat,manuscript=article]{achemso}
\usepackage{chemformula} 
\usepackage[T1]{fontenc} 

\usepackage{float}
\usepackage{graphicx}
\usepackage{pdfpages}
\usepackage{dcolumn}
\usepackage{bm}
\usepackage{mathrsfs,amssymb,amsmath}
\usepackage{ulem}
\usepackage{caption}
\usepackage[colorlinks=true,linkcolor=blue,citecolor=blue,anchorcolor=blue,urlcolor=blue]{hyperref}
\usepackage{booktabs}

\usepackage{setspace}
\usepackage{lineno}
\usepackage[figuresright]{rotating}
\usepackage{pdfpages}

\author{Weiwei He}
\affiliation{State Key Laboratory of Mechanics and Control for Aerospace Structures \& Key Lab for Intelligent Nano Materials and Devices of Ministry of Education \& Institute for Frontier Science, Nanjing University of Aeronautics and Astronautics (NUAA), Nanjing 210016, China}
\altaffiliation{Authors contributed equally.}

\author{Ziming Tang}
\affiliation{State Key Laboratory of Mechanics and Control for Aerospace Structures \& Key Lab for Intelligent Nano Materials and Devices of Ministry of Education \& Institute for Frontier Science, Nanjing University of Aeronautics and Astronautics (NUAA), Nanjing 210016, China}
\altaffiliation{Authors contributed equally.}

\author{Qihua Gong}
\affiliation{State Key Laboratory of Mechanics and Control for Aerospace Structures \& Key Lab for Intelligent Nano Materials and Devices of Ministry of Education \& Institute for Frontier Science, Nanjing University of Aeronautics and Astronautics (NUAA), Nanjing 210016, China}
\alsoaffiliation{MIIT Key Laboratory of Aerospace Information Materials and Physics \& College of Physics, Nanjing University of Aeronautics and Astronautics (NUAA), Nanjing 211106, China}
\email{gongqihua@nuaa.edu.cn}

\author{Min Yi}
\affiliation{State Key Laboratory of Mechanics and Control for Aerospace Structures \& Key Lab for Intelligent Nano Materials and Devices of Ministry of Education \& Institute for Frontier Science, Nanjing University of Aeronautics and Astronautics (NUAA), Nanjing 210016, China}
\email{yimin@nuaa.edu.cn}

\author{Wanlin Guo}
\affiliation{State Key Laboratory of Mechanics and Control for Aerospace Structures \& Key Lab for Intelligent Nano Materials and Devices of Ministry of Education \& Institute for Frontier Science, Nanjing University of Aeronautics and Astronautics (NUAA), Nanjing 210016, China}

\title[AnTitle]{Magnetocaloric effect and its electric-field regulation in CrI$_3$/metal heterostructure}

\abbreviations{IR,NMR,UV}

\begin{document} 
\begin{spacing}{1.25}   
\begin{abstract}
The extraordinary properties of a heterostructure by stacking atom-thick van der Waals (vdW) magnets have been extensively studied. However, the magnetocaloric effect (MCE) of heterostructures that are based on monolayer magnets remains to be explored. Herein, we deliberate MCE of vdW heterostructure composed of a monolayer CrI$_3$ and metal atomic layers (Ag, Hf, Au, and Pb). 
It is revealed that heterostructure engineering by introducing metal substrate can improve MCE of CrI$_3$, particularly boosting relative cooling power to 471.72\,\textmu J\,m$^{-2}$ and adiabatic temperature change to 2.1\,K at 5\,T for CrI$_3$/Hf. This improved MCE is ascribed to the enhancement of magnetic moment and intralayer exchange coupling in CrI$_3$ due to the CrI$_3$/metal heterointerface induced charge transfer. Electric field is further found to tune MCE of CrI$_3$ in heterostructures and could shift the peak temperature by around 10\,K in CrI$_3$/Hf, thus manipulating the working temperature window of MCE. The discovered electric-field and substrate regulated MCE in CrI$_3$/metal heterostructure opens new avenues for low-dimensional magnetic refrigeration.
\end{abstract}


\section{1. INTRODUCTION}

Magnetic cooling based on magnetocaloric effect (MCE) has emerged as a promising alternative to gaseous cooling with its ever-increasing energy consumption and greenhouse gas emissions~\cite{Gutfleisch2011Magnetic,Gottschall2019Making,Kitanovski2020Energy,Hou2022Materials}. 
As the inherited characteristics of magnetic materials, MCE is closely related to the magnetic properties and magnetization behavior under a magnetic field. The unpaired spins in magnetic materials are aligned when a magnetic field is applied, leading to a decrease in magnetic entropy and subsequent release of heat into the surroundings. 
Using MCE, it is possible to achieve target temperatures ranging from ultra-low to room temperature~\cite{Hashimoto1981Magnetic,Zarkevich2020Viable}. 
Up to date, the majority of literature on MCE has focused on bulk magnetocaloric materials~\cite{Pecharsky1997Giant,Luo2006Magnetocaloric,Zhang2010Magnetocaloric,Franco2018Magnetocaloric} or thin films with nanometers to micrometers thickness~\cite{Moya2012Giant,Yuzuak2013Inverse,Khovaylo2014Magnetocaloric}, restricting the application of magnetic refrigeration in compact and miniaturized nanodevices.

In recent years, two-dimensional (2D) van der Waals (vdW) magnets with a wide variety of unconventional properties, which differ from their bulk counterparts~\cite{Huang2017Layer,Gong2017Discovery,Bonilla2018Strong,Deng2018Gate,
Yin2021High,Yin2022Emerging}, have drawn immense interest in fundamental research and device applications~\cite{Li2019Intrinsic,Hou20192D,Wang2020Electrically,Xue2022Nonlinear,
Tang2023Spin}. Layered vdW magnets are bonded to each other through weak vdW forces, allowing the easy separation of monolayers.
For instance, in the monolayer CrI$_3$ that is fabricated by micromechanical exfoliation of bulk CrI$_3$ crystals, the magnetic anisotropy cancels out the thermal fluctuations at finite temperatures and the long-range magnetic order remains~\cite{Huang2017Layer}. Therefore, the spontaneous magnetization in 2D magnets offers new perspectives for exploring MCE down to the monolayer limit.

The application and characterization of monolayer CrI$_3$ usually requires that it be stacked on top of a substrate rather than isolated~\cite{Li2019Pressure,Ubrig2019Low,Guo2021Structural}. 
Thus, vdW magnetic heterostructures with intrinsic magnetism and excellent stacking capability have attracted extensive research~\cite{Zhang2018Strong,Song2018Giant,Gibertini2019Magnetic,Li2020Spin,
Gong2020Electric,Yao2021Recent}. 
Diverse combination of heterostructures formed by CrI$_3$ is an effective approach toward achieving novel properties, such as thermal spin-filtering effect~\cite{Tan2021Spin}, quantum anomalous Hall effect~\cite{Petrov2019Cr,Gao2022Prediction}, half-metallicity~\cite{Zhao2019Nonvolatile,Chakraborty2021Substrate,Wang2021Heterointerface}
, and enhanced magnetic properties~\cite{Chen2019Boosting,Hu2021Exploring,Yu2022Strain}. 
The effect of various substrates on CrI$_3$ in terms of MCE, however, remains to be elucidated, which has stimulated our great interests in MCE of CrI$_3$ in hererostructures.

In the present work, we investigate MCE of CrI$_3$/metal heterostructure and its electric-field ($E$) control. Four types of metals, i.e., Ag, Hf, Au, and Pb, are chosen to construct the heterostructures with monolayer CrI$_3$.
First-principles calculations indicate that metal substrates significantly enhance the magnetic moment and nearest-neighbor exchange interaction of CrI$_3$, while they have different roles in magnetocrystalline energy (MAE).
Atomistic spin simulations show that Curie temperature~($T_{\text{C}}$) and saturation magnetization of CrI$_3$ in heterostructures are boosted compared with the free-standing one. 
In addition, magnetocaloric thermodynamics confirms that heterostructure engineering improves MCE of monolayer CrI$_3$, which is further enhanced in CrI$_3$/Hf by applying a negative $E$. 
Our work not only demystifies tunable MCE in 2D magnets via substrates and $E$, but also opens new vistas for low-dimensional magnetically cooling devices.

\section{2. RESULTS AND DISCUSSION}
\subsection{2.1~Magnetic properties and its electric-field tunability}
\begin{figure*}[!b]
  \includegraphics[width=16cm]{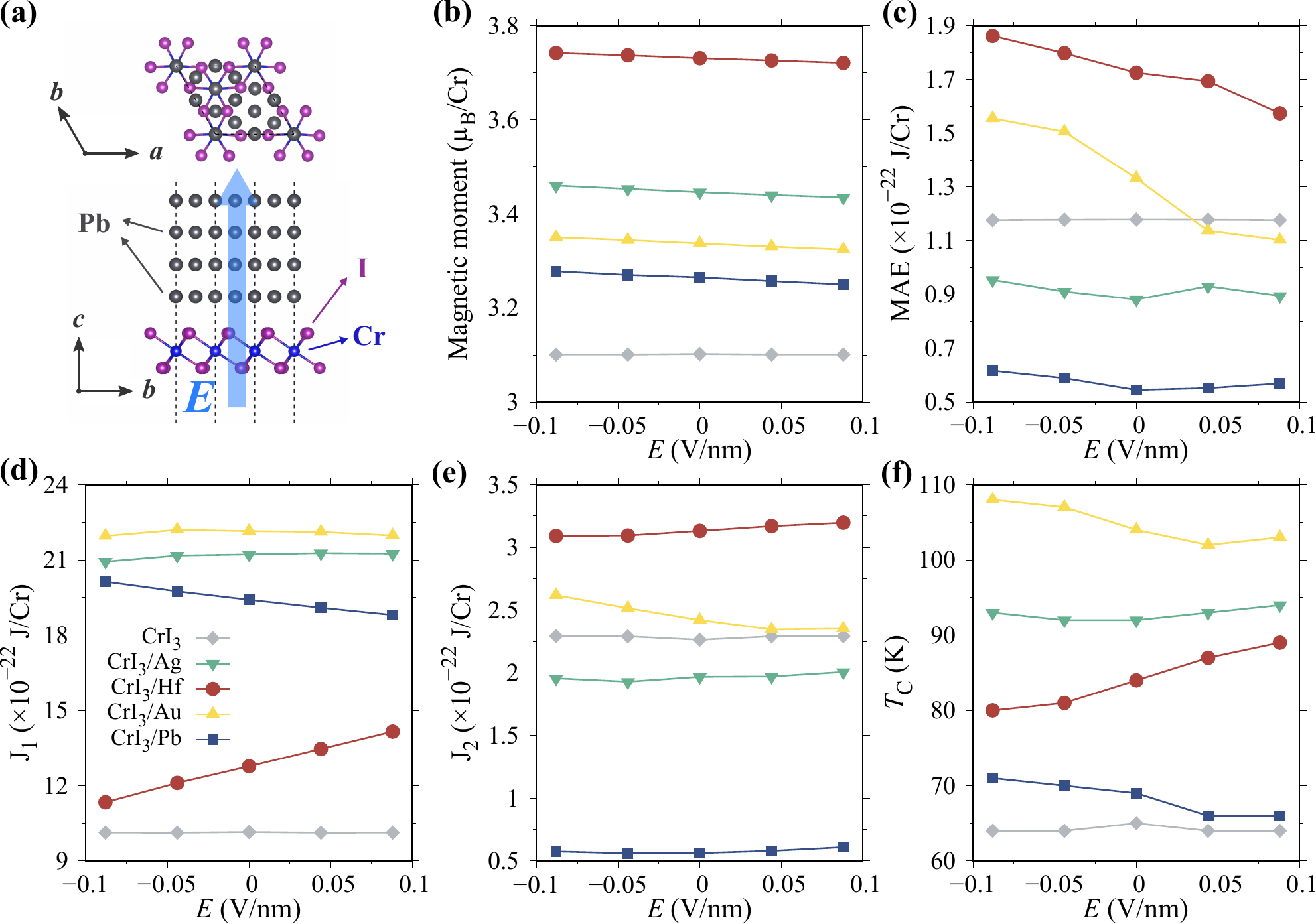}
  \caption{The atomic structure and magnetic properties of CrI$_3$ as a function of $E$ in CrI$_3$/metal vdW heterostructures. (a)\,Top and side views of CrI$_3$/Pb(111) heterostructure. Dark grey, purple, and purplish-blue balls represent Pb, I, and Cr atoms, respectively. (b)\,Magnetic moment per Cr atom, (c)\,MAE, (d) and (e)\,magnetic exchange interaction parameters $J_i$, and (f)\,$T_{\text{C}}$ of CrI$_3$ in monolayer and CrI$_3$/metal heterostructures as a function of $E$.}
  \label{fig:structure}
\end{figure*}

Due to the interfacial interaction, 2D vdW magnets can change their magnetic properties upon contact with other materials. Based on thermodynamic stability assessment, Yang et al. investigated the magnetic properties of CrI$_3$ with 3$d$ transition-metal atoms (from Sc to Zn) absorbed on its surface~\cite{Yang2021Enhancing}. 
In our work, four types of metal are selected (e.g., Ag, Hf, Au, Pb) to construct heterostructures with a favorable lattice match. 
Fig.~\ref{fig:structure}(a) shows the atomic stacking in CrI$_3$/Pb heterostructure. The other three types of CrI$_3$/metal (Ag, Au, and Hf) heterostructure are presented in Fig.~S1 (Supporting Information).
It is noteworthy that CrI$_3$ in heterostructures has fixed lattice parameters of monolayer CrI$_3$, allowing the metals to be compressed or stretched within an acceptable strain range around $\pm$ 5\%.
As illustrated in Fig.~\ref{fig:structure}(a), we are able to investigate $E$ controlled magnetic properties of CrI$_3$/metal heterostructures with the help of dipole layer method.

Magnetic properties of CrI$_3$ are affected by metal substrate and $E$.
As shown in Fig.~\ref{fig:structure}(b)-(f), the impact of different metal substrates on CrI$_3$ are dissimilar. The magnetic moment, $J_1$, and $T_{\text{C}}$ of CrI$_3$ absorbed upon four kinds of metal layers are increased when compared to those of monolayer CrI$_3$. In contrast, MAE of CrI$_3$ is weakened by Ag and Pb substrates, while Au and Hf substrates improve it. The change of $J_2$ shows a similar trend.
Taking CrI$_3$/Hf heterostructure as an example, a detail analysis of the influence of metal substrate on the magnetic moment and MAE of CrI$_3$ is presented.
As can be seen from Fig.~S2, the introduction of Hf intrigues the redistribution of electrons from spin-down channel to spin-up channel on the CrI$_3$ side, thereby increasing the magnetic moment of Cr in Fig.~\ref{fig:structure}(b).
Considering that I atom (5$p$-element) has a much stronger spin-orbit coupling (SOC) effect than Cr (3$d$-element)~\cite{Jiang2018Spin}, $p$-orbital resolved MAE of I is analyzed in Fig.~S3. The enhancement of matrix element differences ($p_x$, $p_y$) compensates for the decrease of ($p_x$, $p_z$), eventually improving MAE of CrI$_3$/Hf in Fig.~\ref{fig:structure}(c).

Magnetic performance of CrI$_3$/metal heterostructures under $E$ exhibit more diverse variations.
Fig.~\ref{fig:structure}(b) shows the effect of $E$ on the magnetic moment of Cr . It can be seen that compared to monolayer CrI$_3$, heterostructures not only achieve an increase in magnetic moment of Cr by promoting spin polarization through charge transfer, but also introduce the strong magneto-electric response that is neglectable in monolayer CrI$_3$.
Under a positive $E$, the magnetic moment of Cr slightly decreases when the strength of $E$ increases. This could be explained by the charge redistribution that leads to a decrease of net charge in the spin-up channel~\cite{Hu2021Exploring, Liu2018Analysis}. Fig.~S2 further shows the charge difference in CrI$_3$/Hf heterostructure under different $E$.
MAE of CrI$_3$ in CrI$_3$/Hf diminishes with the increasing $E$, as shown in Fig.~\ref{fig:structure}(c). Similar to the CrI$_3$/metal heterostructure without $E$, the synergistic effect of matrix element differences ($p_x$, $p_y$) and ($p_x$, $p_z$) promotes the response of MAE to $E$ (Fig.~S3).
Fig.~\ref{fig:structure}(d) and \ref{fig:structure}(e) reveal the trends of ferromagnetic exchange interaction parameters $J_1$ and $J_2$ as a function of $E$. 
From our calculations, it is evident that when $E$ is increased, $J_1$ and $J_2$ of CrI$_3$ in CrI$_3$/Hf are enhanced, while $J_1$ of CrI$_3$/Pb and $J_2$ of CrI$_3$/Au are decreased. To explain the modulation of $J_i$ by $E$ and metal substrate, the energy differences between ferromagnetic (FM) and antiferromagnetic (AFM) states are presented in Fig.~S4.
Similar trends are also found in Fig.~\ref{fig:structure}(f) : $T_{\text{C}}$ of CrI$_3$ in CrI$_3$/Hf increases by about 10\,K as $E$ is raised, whereas $T_{\text{C}}$ of CrI$_3$ adsorbed on Pb and Au decreases by 5\,K.

\subsection{2.2~Demagnetization behavior of CrI$_3$ in heterostructures}
\begin{figure*}[!b]
  \includegraphics[width=16cm]{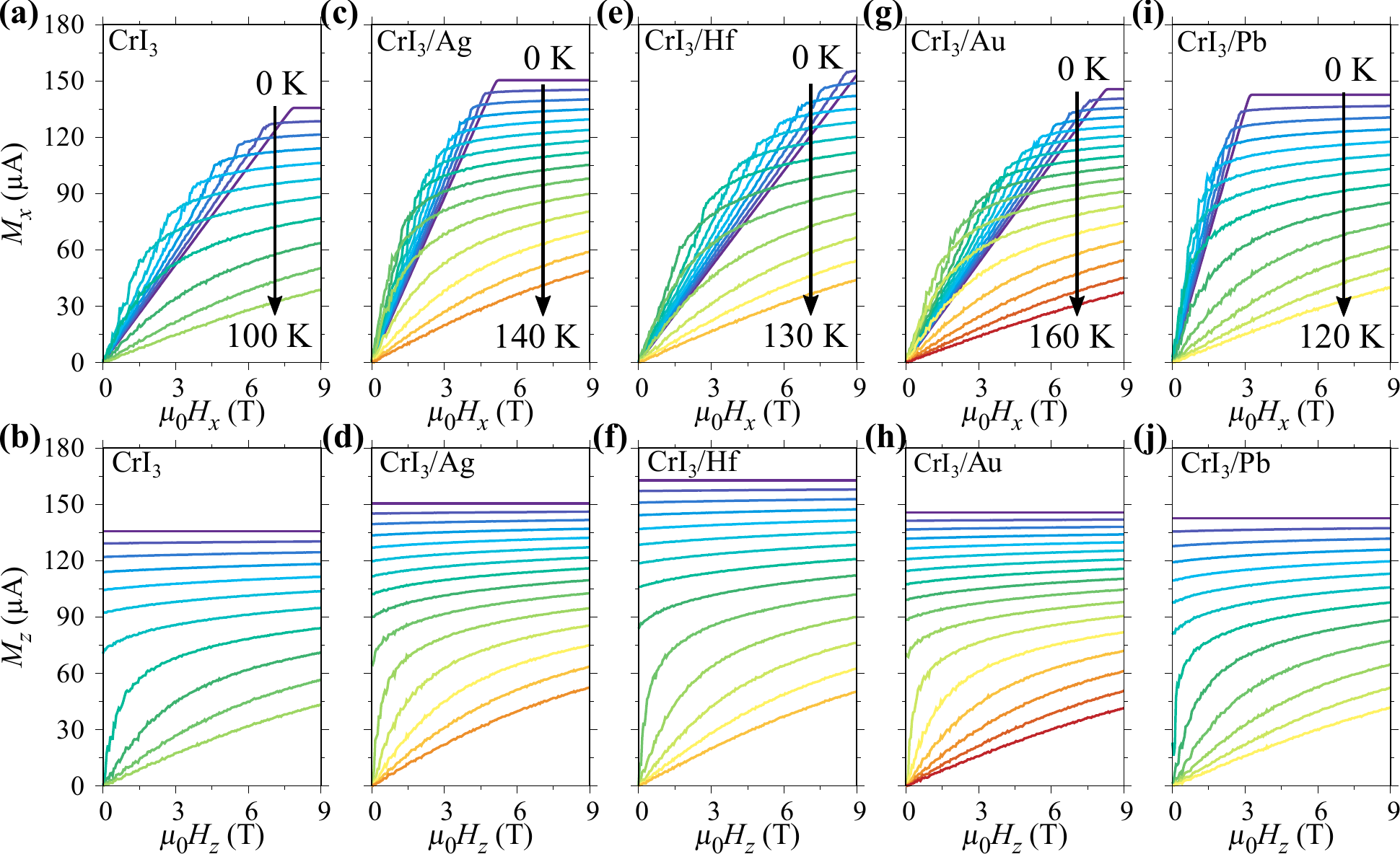}
  \caption{Isothermal magnetization curves of CrI$_3$ in (a,b)\,monolayer CrI$_3$, (c,d)\,CrI$_3$/Ag heterostructure, (e,f)\,CrI$_3$/Hf heterostructure, (g,h)\,CrI$_3$/Au heterostructure, and (i,j)\,CrI$_3$/Pb heterostructure with magnetic field up to 9\,T. The field directions are applied along in-plane for the top-row subfigures and out-of-plane for the bottom-row subfigures. The curves are displayed every 10\,K.}
  \label{fig:M-H}
\end{figure*}

In our previous work~\cite{He2023Giant}, we found that MCE indeed remains in various 2D magnets and can be remarkably tuned by strain.
As a starting point for assessing MCE of CrI$_3$ in heterostructures, we first examine the demagnetization behavior of CrI$_3$ at different temperatures.
Fig.~\ref{fig:M-H} gives the isothermal magnetization curves in monolayer CrI$_3$ and heterostructures under in-plane and out-of-plane $H$. The magnetization vector of Cr atoms turns to the direction of applied field as $H$ increases, and gradually becomes saturated under a high field. Moreover, saturation magnetization decreases with increasing temperature. 
Meanwhile, saturation magnetization of CrI$_3$ at 0\,K should be proportional to the magnetic moment of Cr atom. Thus, CrI$_3$/Hf heterostructure at the same temperature is expected to have the highest saturation magnetization according to Fig.~\ref{fig:structure}(b).

Figure~\ref{fig:M-H} also confirms the anisotropy of demagnetization behavior. The out-of-plane magnetic fields saturate magnetization curves much more easily than in-plane ones, owing to the out-of-plane MAE in Fig.~\ref{fig:structure}(c). 
This anisotropic phenomenon has been explained at length in bulk CrI$_3$~\cite{Liu2018Anisotropic,Tran2022Insight}, and conclusions are also applicable to monolayers.
When compared to monolayer CrI$_3$, the magnetization curves of CrI$_3$ in CrI$_3$/Ag and CrI$_3$/Pb heterostructures tend to saturation at lower fields (Fig.~\ref{fig:M-H}(c) and(i)), while those of CrI$_3$ in CrI$_3$/Hf and CrI$_3$/Au heterostructures remain unsaturated under $H$ up to 8\,T (Fig.~\ref{fig:M-H}(e) and(d)). This is related to different effects of metal substrate on MAE of CrI$_3$, which is similar to the trend in Fig.~\ref{fig:structure}(c).
In addition, magnetization curves in different heterostructures are distinguished at the same temperature. Compared with the magnetization curve of monolayer CrI$_3$ at 100\,K, CrI$_3$ in CrI$_3$/Au shows the same magnetization degree until 160\,K.
Magnetization curves of CrI$_3$ in CrI$_3$/Hf at different $E$ are supplemented in Fig.~S5. The saturation magnetization at 0\,K and the anisotropy under different $E$ only change slightly under $E$ modulation.

\subsection{2.3~Magnetocaloric effect and its electric-field tunability}

\begin{figure*}[!b]
  \includegraphics[width=16cm]{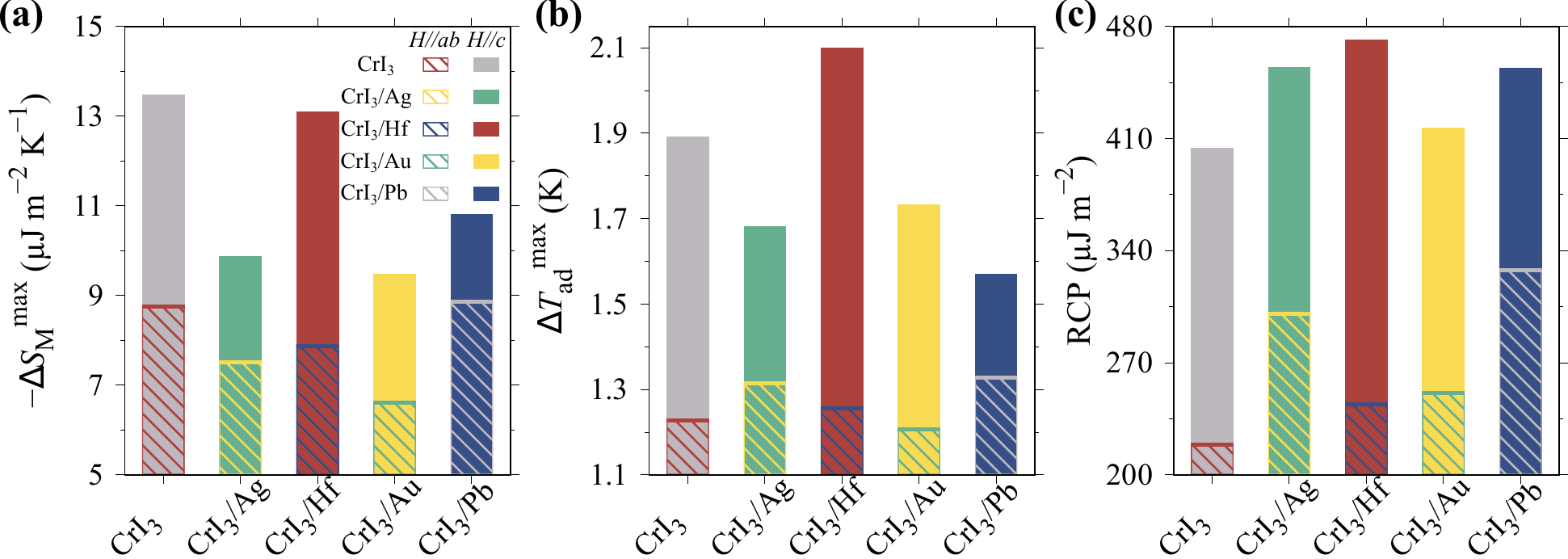}
  \caption{Influence of metal substrates on MCE of CrI$_3$ under a magnetic field of 5\,T applied in different directions. (a)\,Maximum magnetic entropy change ($-\Delta S_\text{M}^\text{max}$), (b)\,maximum adiabatic temperature change ($\Delta T_\text{ad}^\text{max}$), and (c)\,relative cooling power (RCP).}
  \label{fig:5T}
\end{figure*}

To explore the influence of metal substrates on MCE, maximum magnetic entropy change ($-\Delta S_\text{M}^\text{max}$), maximum adiabatic temperature change ($\Delta T_\text{ad}^\text{max}$), and relative cooling power (RCP) with a magnetic field of 5\,T applied in different directions are shown in Fig.~\ref{fig:5T}. 
$-\Delta S_\text{M}^\text{max}$, $\Delta T_\text{ad}^\text{max}$, and RCP of monolayer CrI$_3$ in our work are calculated to be 13.48\,\textmu J\,m$^{-2}$\,K, 1.89\,K, and 404.35\,\textmu J\,m$^{-2}$, respectively. The estimated results of monolayer CrI$_3$ are in good agreement with the experimental work of bulk counterpart~\cite{Liu2018Anisotropic}, whose $-\Delta S_\text{M}^\text{max}$ and $\Delta T_\text{ad}^\text{max}$ are measured as around 3.8\,J\,kg$^{-1}$\,K and 1.55\,K  under an out-of-plane magnetic field of 5\,T.
It also can be seen in Fig.~\ref{fig:5T} that metal substrates do improve MCE of CrI$_3$, particularly in terms of increasing RCP.
CrI$_3$ in all heterostructures achieves a higher RCP compared with monolayer CrI$_3$, as shown in Fig.~\ref{fig:5T}(c).
Among them, RCP of CrI$_3$ in CrI$_3$/Hf heterostructure is as high as 471.72\,\textmu J\,m$^{-2}$ at 5\,T, suggesting there is more heat transferred between hot and cold reservoirs during a magnetic refrigeration cycle.
In heterostructures, however, $-\Delta S_\text{M}^\text{max}$ of CrI$_3$ absorbed on four types of metal substrates does not surpass that of monolayer CrI$_3$ (Fig.~\ref{fig:5T}(a)), mainly owing to the improved $T_\text{C}$ of CrI$_3$. More specifically, according to Eq~\ref{eq:S}, materials with superior $-\Delta S_\text{M}^\text{max}$ possess a comparatively larger magnetization and a relatively low $T_\text{C}$. The small increase in the magnetization cannot counteract the substantial improvement in $T_\text{C}$ of CrI$_3$ in heterostructures, thus leading to a decrease in $-\Delta S_\text{M}^\text{max}$.
In contrast to $-\Delta S_\text{M}^\text{max}$, $\Delta T_\text{ad}^\text{max}$ reflects the enhancement of MCE by metal substrates.
On account of the significantly improved magnetic moment and moderate $T_\text{C}$, Hf substrate can increase $\Delta T_\text{ad}^\text{max}$ of CrI$_3$ by 10.7\% (up to 2.1\,K) under an out-of-plane magnetic field, as shown in Fig.~\ref{fig:5T}(b). 
The correlation between MCE and the applied field directions is also present in Fig.~\ref{fig:5T}. It is clear that the difference between in-plane and out-of-plane MCE in CrI$_3$/Hf heterostructure is larger than that in monolayer CrI$_3$, while it is smaller in CrI$_3$/Ag and CrI$_3$/Pb heterostructures.
The effect of substrates on anisotropic MCE in CrI$_3$ is also essentially attributed to the effect of substrates on MAE (Fig.~\ref{fig:structure}(c)).

\begin{figure*}[!b]
  \includegraphics[width=16cm]{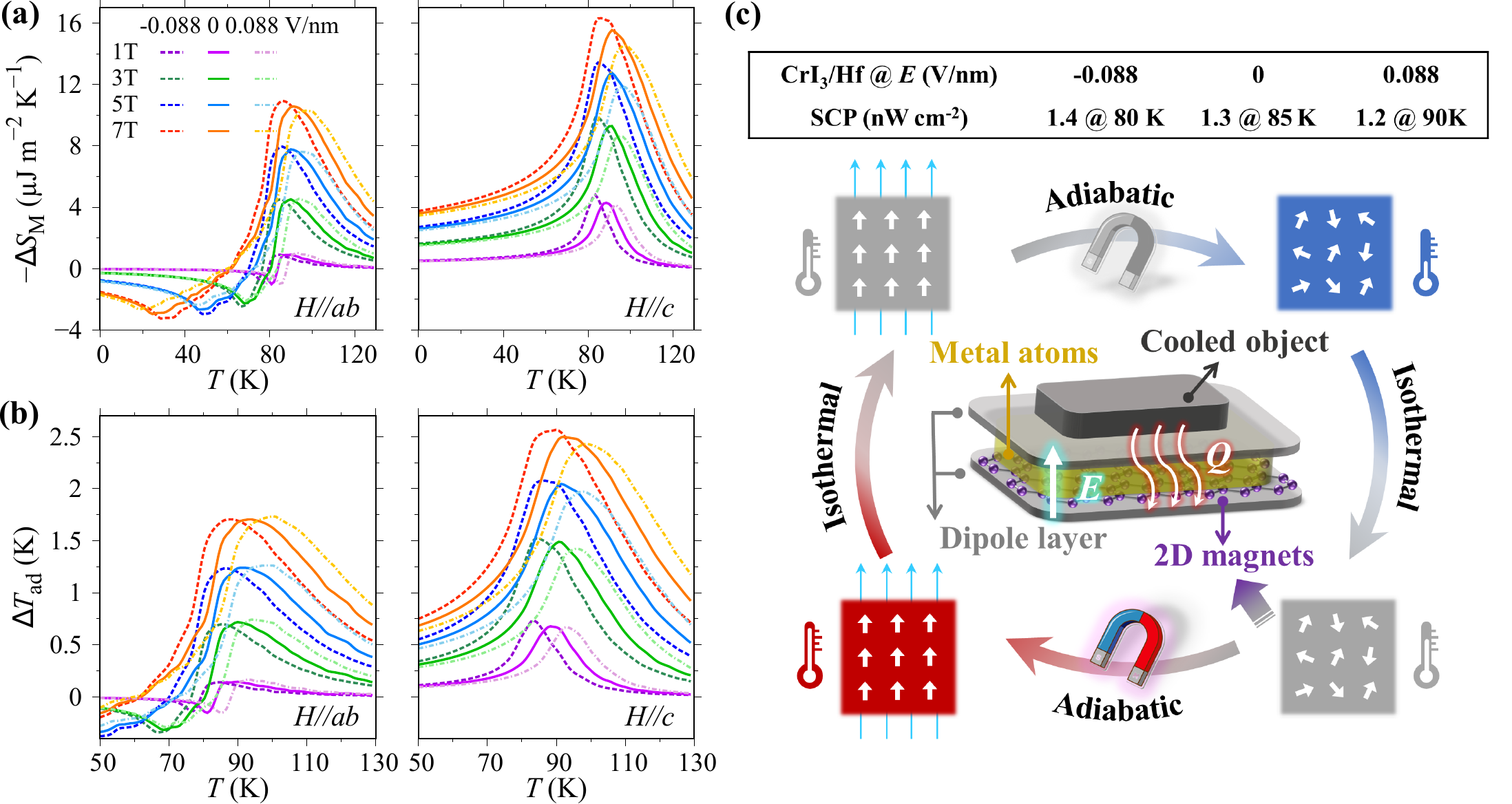}
  \caption{Electric-field-tunable MCE and its potential application in magnetic cooling. $E$-tunable (a)\,isothermal magnetic entropy change ($-\Delta S_\text{M}$) and (b)\,adiabatic temperature change ($\Delta T_\text{ad}$) \textit{vs} $T$ curves of CrI$_3$ in CrI$_3$/Hf heterostructure. (c)\,Schematic illustration of Carnot refrigeration cycle using MCE of CrI$_3$/metal heterostructure under $E$, in which specific cooling power (SCP) of CrI$_3$/Hf heterostructure at 5\,T is estimated.}
  \label{fig:E}
\end{figure*}

Figure~\ref{fig:E}(a) and (b) present the $E$-tunable MCE of CrI$_3$ in CrI$_3$/Hf heterostructure under the in-plane and out-of-plane $H$. 
Our calculation results show that, the magnetocaloric performance of CrI$_3$ in CrI$_3$/Hf heterostructure is further enhanced at a negative $E$, while the isothermal magnetic entropy change ($-\Delta S_\text{M}$) decreased slightly at a positive $E$ (Fig.~\ref{fig:E}(a)). 
This can be understood from Eq.\ref{eq:S}, which indicates that the partial derivative of magnetization with respect to temperature determines the entropy change. As shown in Fig.~S5, a negative $E$ induces the magnetization of CrI$_3$ in CrI$_3$/Hf to be more sensitive to temperature changes than a positive $E$, resulting in higher $-\Delta S_\text{M}^\text{max}$ under the same $H$. A similar explanation can be used for $\Delta T_\text{ad}^\text{max}$ in Fig.~\ref{fig:E}(b).
In addition, there is a consistent trend between the shift of peak temperature position in MCE curves and the adjustment of $T_\text{C}$ regulated by $E$, as shown in Fig.~\ref{fig:E}(a) and (b). By applying $E$ to CrI$_3$/Hf heterostructure, the peak temperature of $-\Delta S_\text{M}$ and adiabatic temperature change ($\Delta T_\text{ad}$) curves is shift up to 10\,K, thereby extending the working temperature window of MCE for CrI$_3$/metal heterostructures.
$-\Delta S_\text{M}$ and $\Delta T_\text{ad}$ curves of CrI$_3$ in other heterostructures are given in Fig.~S6 and Fig.~S7, and these curves are almost unaffected by $E$ modulation.

Schematic illustration of magnetic Carnot refrigeration cycle using CrI$_3$/metal heterostructure under $E$ is shown in Fig.~\ref{fig:E}(c). We propose an ideal assumption that when the cooled object touches a heterostructure under an applied periodic magnetic field, heat can be completely transferred outward from the cooled object. With an operating frequency of 1\,Hz, the ideal specific cooling power (SCP) of CrI$_3$/Hf heterostructure around 85\,K is close to 1.3\,nW\,cm$^2$ at 5\,T.  Compared to monolayer CrI$_3$~\cite{He2023Giant}, CrI$_3$/Hf heterostructure has an operating temperature increase of roughly 20\,K owing to the metal substrate, and the operating temperature could be further shift up to 10\,K by applying $E$.

\section{3. CONCLUSIONS}
In summary, MCE of monolayer CrI$_3$ absorbed on various metal substrates and its $E$ modulation is comprehensively investigated using first-principles calculations, atomistic spin simulations, and magnetocaloric thermodynamics.
Magnetic properties of CrI$_3$ are affected by metal substrates and exhibit more diverse variations under $E$, mainly owing to the charge transfer at the interface.
Ag, Hf, Au, and Pb substrate are demonstrated to improve the magnetocaloric performance of CrI$_3$, particularly with RCP and $\Delta T_\text{ad}$ of CrI$_3$ in CrI$_3$/Hf are increased to 471.72\,\textmu J\,m$^{-2}$ and 2.1\,K at 5\,T, respectively.
Ag and Pb are found to weaken the MCE anisotropy, while Hf enhances it. 
Moreover, CrI$_3$ in CrI$_3$/Hf with a negative $E$ exhibits better magnetocaloric performance than that with a positive one. Through $E$ modulation, the peak temperature of $-\Delta S_\text{M}$ and $\Delta T_\text{ad}$ curves in CrI$_3$/Hf can be shift up to 10\,K, allowing for widely tunable working temperature window of MCE.
Our results on MCE of heterostructure formed by monolayer magnets and its $E$ regulation provide an important basis for designing and building authentic low-dimensional magnetic refrigeration devices.

\section{4. METHODOLOGY}
The first-principles calculations within density functional theory are carried out to calculate the magnetic properties of monolayer CrI$_3$ absorbed on four types of metal substrates (Ag, Hf, Au, and Pb) by using \textit{Vienna Ab initio Simulation Package}~(VASP)~\cite{Kresse1996Efficient,Kresse1999From}. The exchange-correlation functional is treated with the the generalized gradient approximation~(GGA) of the Perdew--Burke--Ernzerhof~(PBE) form~\cite{Perdew1996Generalized}. The CrI$_3$/metal heterostructures are constructed by stacking four metallic layers onto a monolayer CrI$_3$ sheet. To avoid the interaction between neighboring slabs, the vacuum space along the $z$-axis is set to 35\,\AA ~\cite{Neugebauer1992Adsorbate}. A cutoff energy of 500\,eV is utilized. The convergence criteria for energy and force in structure relaxation are 10$^{-5}$\,eV and 0.01\,eV/\AA , respectively. The energy convergence is 10$^{-6}$\,eV in self-consistent electronic calculations. The $5\times5\times1$, $11\times11\times1$, and $11\times11\times1$ Monkhorst--Pack $k$-mesh in the CrI$_3$/Pb heterostructure ($3\times3\times1$, $5\times5\times1$, and $5\times5\times1$ in other heterostructures that contain more atoms) are adopted in the ionic optimization, electronic optimization, and MAE calculation, respectively~\cite{Monkhorst1976Special}. The MAE is obtained by calculating the energy differences between the spin quantization axes whose directions are aligned with different crystallographic axes. In order to calculate the MAE, SOC is considered\cite{Vaz2008Magnetism}. The magnetic exchange parameters are determined by substituting the magnetic configuration energies into the classic spin Hamiltonian~\cite{Skubic2008A}
\begin{equation}
\begin{aligned}
H &= E_0 - \frac{1}{2} J_1\sum_N \mathbf s_i \cdot \mathbf s_j - \frac{1}{2} J_2\sum_{NN} \mathbf s_i \cdot \mathbf s_j  ,
\end{aligned}
\label{eq:H}
\end{equation}
where $E_0$ is the energy without spin contribution, $\mathbf s_i$ represents the unit vector of the atomistic spin direction at atom $i$. $J_1$ and $J_2$ denote the nearest-neighbour~(NN) and next-NN exchange interaction parameters, respectively.

After obtaining the magnetic exchange parameters from first-principles calculations, 
$T_{\text{C}}$ and temperature-dependent magnetization can be determined by the atomic spin model that has been numerically implemented in VAMPIRE~\cite{Evans2014Atomistic,Gong2019Calculating,Gong2019Multiscale}. The demagnetization field caused by the atomistic spins themselves is also considered. The figures of merit for MCE are generally described by $\Delta S_\text{M}$ and $\Delta T_\text{ad}$ upon a variation of magnetic field~($H$). Based on the classical thermodynamics and the Maxwell relation, $\Delta S_\text{M}$ is given by~\cite{Pecharsky1999Magnetocaloric,Gschneidner2000Magnetocaloric}
\begin{equation}
\Delta S_\text{M} = \int_0^H\left(\frac{\partial S}{\partial H}\right)_T\text{d}H = \mu_0\int_0^H\left(\frac{\partial M}{\partial T}\right)_H\text{d}H,
\label{eq:S}
\end{equation}
where $S$ and $M$ refer to entropy and magnetization, respectively. $\mu_0$ is vacuum permeability. Normally, MCE is characterized by $-\Delta S_\text{M}$, given that the degree of disorder in the magnetic moment decreases with increasing $H$. $\Delta T_\text{ad}$ can be similarly calculated as
\begin{equation}
\begin{aligned}
\Delta T_\text{ad} &= -\mu_0\int_0^H \frac{T}{\rho c_\text{p}}\left(\frac{\partial S}{\partial H}\right)_{T}\text{d}H \\
&= -\mu_0\int_0^H \frac{T}{\rho c_\text{p}}\left(\frac{\partial M}{\partial T}\right)_{H}\text{d}H,
\end{aligned}
\label{eq:T}
\end{equation}
where $\rho$ and $c_\text{p}$ are density and specific heat capacity of monolayer CrI$_3$, respectively. RCP as another descriptor for MCE is used to characterize the heat transfer across reservoirs and reveal the potential MCE in magnets, which is calculated as
\begin{equation}
\label{eq:RCP}
\text{RCP}=\left | \Delta S_\text{M}^\text{max} \right |  \times \delta T_\text{FWHM},
\end{equation}
where the $\delta T_\text{FWHM}$ means the full width at half maximum of the $-\Delta S_\text{M}$ \textit{vs} $T$ curve.

\section*{Supporting Information}

Top and side views of CrI$_3$/metal vdW heterostructures; Spin-dependent plane integrated charge density difference along $c/z$ direction; The $p$-orbital resolved MAE of I atom in CrI$_3$; Energy difference between FM and AFM configurations of CrI$_3$ in CrI$_3$/metal vdW heterostructures; Isothermal demagnetization curves of CrI$_3$ in CrI$_3$/Hf heterostructure under different $E$; Electric-field-tunable $-\Delta S_\text{M}$ \textit{vs} $T$ curves for CrI$_3$/metal heterostructures; Electric-field-tunable $\Delta T_\text{ad}$ \textit{vs} $T$ curves for CrI$_3$/metal heterostructures.

\section*{Acknowledgment}
The authors acknowledge the support from the National Natural Science Foundation of China (12272173, 11902150), the National Overseas Thousand Youth Talents Program, the Research Fund of State Key Laboratory of Mechanics and Control of Mechanical Structures (MCMS-I-0419G01 and MCMS-I-0421K01), a project Funded by the Priority Academic Program Development of Jiangsu Higher Education Institutions, and the Interdisciplinary Innovation Fund for Doctoral Students of Nanjing University of Aeronautics and Astronautics (KXKCXJJ202306). This work is partially supported by High Performance Computing Platform of Nanjing University of Aeronautics and Astronautics. Simulations were also performed on Hefei advanced computing center.
\normalem
\bibliography{heterostructure}

\providecommand{\latin}[1]{#1}
\makeatletter
\providecommand{\doi}
  {\begingroup\let\do\@makeother\dospecials
  \catcode`\{=1 \catcode`\}=2 \doi@aux}
\providecommand{\doi@aux}[1]{\endgroup\texttt{#1}}
\makeatother
\providecommand*\mcitethebibliography{\thebibliography}
\csname @ifundefined\endcsname{endmcitethebibliography}
  {\let\endmcitethebibliography\endthebibliography}{}
\begin{mcitethebibliography}{61}
\providecommand*\natexlab[1]{#1}
\providecommand*\mciteSetBstSublistMode[1]{}
\providecommand*\mciteSetBstMaxWidthForm[2]{}
\providecommand*\mciteBstWouldAddEndPuncttrue
  {\def\EndOfBibitem{\unskip.}}
\providecommand*\mciteBstWouldAddEndPunctfalse
  {\let\EndOfBibitem\relax}
\providecommand*\mciteSetBstMidEndSepPunct[3]{}
\providecommand*\mciteSetBstSublistLabelBeginEnd[3]{}
\providecommand*\EndOfBibitem{}
\mciteSetBstSublistMode{f}
\mciteSetBstMaxWidthForm{subitem}{(\alph{mcitesubitemcount})}
\mciteSetBstSublistLabelBeginEnd
  {\mcitemaxwidthsubitemform\space}
  {\relax}
  {\relax}

\bibitem[Gutfleisch \latin{et~al.}(2011)Gutfleisch, Willard, Br{\"{u}}ck, Chen,
  Sankar, and Liu]{Gutfleisch2011Magnetic}
Gutfleisch,~O.; Willard,~M.~A.; Br{\"{u}}ck,~E.; Chen,~C.~H.; Sankar,~S.~G.;
  Liu,~J.~P. {Magnetic materials and devices for the 21st century: Stronger,
  lighter, and more energy efficient}. \emph{Advanced Materials} \textbf{2011},
  \emph{23}, 821--842\relax
\mciteBstWouldAddEndPuncttrue
\mciteSetBstMidEndSepPunct{\mcitedefaultmidpunct}
{\mcitedefaultendpunct}{\mcitedefaultseppunct}\relax
\EndOfBibitem
\bibitem[Gottschall \latin{et~al.}(2019)Gottschall, Skokov, Fries, Taubel,
  Radulov, Scheibel, Benke, Riegg, and Gutfleisch]{Gottschall2019Making}
Gottschall,~T.; Skokov,~K.~P.; Fries,~M.; Taubel,~A.; Radulov,~I.;
  Scheibel,~F.; Benke,~D.; Riegg,~S.; Gutfleisch,~O. {Making a cool choice: The
  materials library of magnetic refrigeration}. \emph{Advanced Energy
  Materials} \textbf{2019}, \emph{9}, 1901322\relax
\mciteBstWouldAddEndPuncttrue
\mciteSetBstMidEndSepPunct{\mcitedefaultmidpunct}
{\mcitedefaultendpunct}{\mcitedefaultseppunct}\relax
\EndOfBibitem
\bibitem[Kitanovski and Kitanovski(2020)Kitanovski, and
  Kitanovski]{Kitanovski2020Energy}
Kitanovski,~A.; Kitanovski,~A. {Energy applications of magnetocaloric
  materials}. \emph{Advanced Energy Materials} \textbf{2020}, \emph{10},
  1903741\relax
\mciteBstWouldAddEndPuncttrue
\mciteSetBstMidEndSepPunct{\mcitedefaultmidpunct}
{\mcitedefaultendpunct}{\mcitedefaultseppunct}\relax
\EndOfBibitem
\bibitem[Hou \latin{et~al.}(2022)Hou, Qian, and Takeuchi]{Hou2022Materials}
Hou,~H.; Qian,~S.; Takeuchi,~I. {Materials, physics and systems for
  multicaloric cooling}. \emph{Nature Reviews Materials} \textbf{2022},
  \emph{7}, 633--652\relax
\mciteBstWouldAddEndPuncttrue
\mciteSetBstMidEndSepPunct{\mcitedefaultmidpunct}
{\mcitedefaultendpunct}{\mcitedefaultseppunct}\relax
\EndOfBibitem
\bibitem[Hashimoto \latin{et~al.}(1981)Hashimoto, Numasawa, Shino, and
  Okada]{Hashimoto1981Magnetic}
Hashimoto,~T.; Numasawa,~T.; Shino,~M.; Okada,~T. {Magnetic refrigeration in
  the temperature range from 10~K to room temperature: The ferromagnetic
  refrigerants}. \emph{Cryogenics} \textbf{1981}, \emph{21}, 647--653\relax
\mciteBstWouldAddEndPuncttrue
\mciteSetBstMidEndSepPunct{\mcitedefaultmidpunct}
{\mcitedefaultendpunct}{\mcitedefaultseppunct}\relax
\EndOfBibitem
\bibitem[Zarkevich and Zverev(2020)Zarkevich, and Zverev]{Zarkevich2020Viable}
Zarkevich,~N.~A.; Zverev,~V.~I. {Viable materials with a giant magnetocaloric
  effect}. \emph{Crystals} \textbf{2020}, \emph{10}, 815\relax
\mciteBstWouldAddEndPuncttrue
\mciteSetBstMidEndSepPunct{\mcitedefaultmidpunct}
{\mcitedefaultendpunct}{\mcitedefaultseppunct}\relax
\EndOfBibitem
\bibitem[Pecharsky and Gschneidner(1997)Pecharsky, and
  Gschneidner]{Pecharsky1997Giant}
Pecharsky,~V.~K.; Gschneidner,~K.~A. {Giant Magnetocaloric Effect
  in$\text{Gd}_\text{5}(\text{Si}_\text{2}\text{Ge}_\text{2})$}. \emph{Physical
  Review Letters} \textbf{1997}, \emph{78}, 4494\relax
\mciteBstWouldAddEndPuncttrue
\mciteSetBstMidEndSepPunct{\mcitedefaultmidpunct}
{\mcitedefaultendpunct}{\mcitedefaultseppunct}\relax
\EndOfBibitem
\bibitem[Luo \latin{et~al.}(2006)Luo, Zhao, Pan, and
  Wang]{Luo2006Magnetocaloric}
Luo,~Q.; Zhao,~D.~Q.; Pan,~M.~X.; Wang,~W.~H. {Magnetocaloric effect in
  Gd-based bulk metallic glasses}. \emph{Applied Physics Letters}
  \textbf{2006}, \emph{89}, 081914\relax
\mciteBstWouldAddEndPuncttrue
\mciteSetBstMidEndSepPunct{\mcitedefaultmidpunct}
{\mcitedefaultendpunct}{\mcitedefaultseppunct}\relax
\EndOfBibitem
\bibitem[Zhang \latin{et~al.}(2010)Zhang, Tejada, Xin, Sun, Wong, and
  Bohigas]{Zhang2010Magnetocaloric}
Zhang,~X.~X.; Tejada,~J.; Xin,~Y.; Sun,~G.~F.; Wong,~K.~W.; Bohigas,~X.
  {Magnetocaloric effect in
  $\text{La}_\text{0.67}\text{Ca}_\text{0.33}\text{MnO}_\text{$\delta$}$ and
  $\text{La}_\text{0.60}\text{Y}_\text{0.07}\text{Ca}_\text{0.33}\text{MnO}_\text{$\delta$}$
  bulk materials}. \emph{Applied Physics Letters} \textbf{2010}, \emph{69},
  3596\relax
\mciteBstWouldAddEndPuncttrue
\mciteSetBstMidEndSepPunct{\mcitedefaultmidpunct}
{\mcitedefaultendpunct}{\mcitedefaultseppunct}\relax
\EndOfBibitem
\bibitem[Franco \latin{et~al.}(2018)Franco, Bl{\'a}zquez, Ipus, Law,
  Moreno-Ram{\'\i}rez, and Conde]{Franco2018Magnetocaloric}
Franco,~V.; Bl{\'a}zquez,~J.; Ipus,~J.; Law,~J.; Moreno-Ram{\'\i}rez,~L.;
  Conde,~A. {Magnetocaloric effect: From materials research to refrigeration
  devices}. \emph{Progress in Materials Science} \textbf{2018}, \emph{93},
  112--232\relax
\mciteBstWouldAddEndPuncttrue
\mciteSetBstMidEndSepPunct{\mcitedefaultmidpunct}
{\mcitedefaultendpunct}{\mcitedefaultseppunct}\relax
\EndOfBibitem
\bibitem[Moya \latin{et~al.}(2012)Moya, Hueso, Maccherozzi, Tovstolytkin,
  Podyalovskii, Ducati, Phillips, Ghidini, Hovorka, Berger, Vickers, Defay,
  Dhesi, and Mathur]{Moya2012Giant}
Moya,~X.; Hueso,~L.~E.; Maccherozzi,~F.; Tovstolytkin,~A.~I.;
  Podyalovskii,~D.~I.; Ducati,~C.; Phillips,~L.~C.; Ghidini,~M.; Hovorka,~O.;
  Berger,~A.; Vickers,~M.~E.; Defay,~E.; Dhesi,~S.~S.; Mathur,~N.~D. {Giant and
  reversible extrinsic magnetocaloric effects in
  $\text{La}_\text{0.7}\text{Ca}_\text{0.3}\text{MnO}_\text{3}$ films due to
  strain}. \emph{Nature Materials} \textbf{2012}, \emph{12}, 52--58\relax
\mciteBstWouldAddEndPuncttrue
\mciteSetBstMidEndSepPunct{\mcitedefaultmidpunct}
{\mcitedefaultendpunct}{\mcitedefaultseppunct}\relax
\EndOfBibitem
\bibitem[Y{\"{u}}z{\"{u}}ak \latin{et~al.}(2013)Y{\"{u}}z{\"{u}}ak, Dincer,
  Elerman, Auge, Teichert, and H{\"{u}}tten]{Yuzuak2013Inverse}
Y{\"{u}}z{\"{u}}ak,~E.; Dincer,~I.; Elerman,~Y.; Auge,~A.; Teichert,~N.;
  H{\"{u}}tten,~A. {Inverse magnetocaloric effect of epitaxial Ni-Mn-Sn thin
  films}. \emph{Applied Physics Letters} \textbf{2013}, \emph{103},
  222403\relax
\mciteBstWouldAddEndPuncttrue
\mciteSetBstMidEndSepPunct{\mcitedefaultmidpunct}
{\mcitedefaultendpunct}{\mcitedefaultseppunct}\relax
\EndOfBibitem
\bibitem[Khovaylo \latin{et~al.}(2014)Khovaylo, Rodionova, Shevyrtalov, and
  Novosad]{Khovaylo2014Magnetocaloric}
Khovaylo,~V.~V.; Rodionova,~V.~V.; Shevyrtalov,~S.~N.; Novosad,~V.
  {Magnetocaloric effect in “reduced” dimensions: Thin films, ribbons, and
  microwires of Heusler alloys and related compounds}. \emph{Physica Status
  Solidi B} \textbf{2014}, \emph{251}, 2104--2113\relax
\mciteBstWouldAddEndPuncttrue
\mciteSetBstMidEndSepPunct{\mcitedefaultmidpunct}
{\mcitedefaultendpunct}{\mcitedefaultseppunct}\relax
\EndOfBibitem
\bibitem[Huang \latin{et~al.}(2017)Huang, Clark, Navarro-Moratalla, Klein,
  Cheng, Seyler, Zhong, Schmidgall, McGuire, Cobden, \latin{et~al.}
  others]{Huang2017Layer}
Huang,~B.; Clark,~G.; Navarro-Moratalla,~E.; Klein,~D.~R.; Cheng,~R.;
  Seyler,~K.~L.; Zhong,~D.; Schmidgall,~E.; McGuire,~M.~A.; Cobden,~D.~H.,
  \latin{et~al.}  {Layer-dependent ferromagnetism in a van der Waals crystal
  down to the monolayer limit}. \emph{Nature} \textbf{2017}, \emph{546},
  270--273\relax
\mciteBstWouldAddEndPuncttrue
\mciteSetBstMidEndSepPunct{\mcitedefaultmidpunct}
{\mcitedefaultendpunct}{\mcitedefaultseppunct}\relax
\EndOfBibitem
\bibitem[Gong \latin{et~al.}(2017)Gong, Li, Li, Ji, Stern, Xia, Cao, Bao, Wang,
  Wang, \latin{et~al.} others]{Gong2017Discovery}
Gong,~C.; Li,~L.; Li,~Z.; Ji,~H.; Stern,~A.; Xia,~Y.; Cao,~T.; Bao,~W.;
  Wang,~C.; Wang,~Y., \latin{et~al.}  {Discovery of intrinsic ferromagnetism in
  two-dimensional van der Waals crystals}. \emph{Nature} \textbf{2017},
  \emph{546}, 265--269\relax
\mciteBstWouldAddEndPuncttrue
\mciteSetBstMidEndSepPunct{\mcitedefaultmidpunct}
{\mcitedefaultendpunct}{\mcitedefaultseppunct}\relax
\EndOfBibitem
\bibitem[Bonilla \latin{et~al.}(2018)Bonilla, Kolekar, Ma, Diaz, Kalappattil,
  Das, Eggers, Gutierrez, Phan, and Batzill]{Bonilla2018Strong}
Bonilla,~M.; Kolekar,~S.; Ma,~Y.; Diaz,~H.~C.; Kalappattil,~V.; Das,~R.;
  Eggers,~T.; Gutierrez,~H.~R.; Phan,~M.~H.; Batzill,~M. {Strong roomerature
  ferromagnetism in $\text{VSe}_{\text{2}}$ monolayers on van der Waals
  substrates}. \emph{Nature Nanotechnology} \textbf{2018}, \emph{13},
  289--293\relax
\mciteBstWouldAddEndPuncttrue
\mciteSetBstMidEndSepPunct{\mcitedefaultmidpunct}
{\mcitedefaultendpunct}{\mcitedefaultseppunct}\relax
\EndOfBibitem
\bibitem[Deng \latin{et~al.}(2018)Deng, Yu, Song, Zhang, Wang, Sun, Yi, Wu, Wu,
  Zhu, Wang, Chen, and Zhang]{Deng2018Gate}
Deng,~Y.; Yu,~Y.; Song,~Y.; Zhang,~J.; Wang,~N.~Z.; Sun,~Z.; Yi,~Y.; Wu,~Y.~Z.;
  Wu,~S.; Zhu,~J.; Wang,~J.; Chen,~X.~H.; Zhang,~Y. {Gate-tunable
  room-temperature ferromagnetism in two-dimensional
  $\text{Fe}_{\text{3}}\text{GeTe}_{\text{2}}$}. \emph{Nature} \textbf{2018},
  \emph{563}, 94--99\relax
\mciteBstWouldAddEndPuncttrue
\mciteSetBstMidEndSepPunct{\mcitedefaultmidpunct}
{\mcitedefaultendpunct}{\mcitedefaultseppunct}\relax
\EndOfBibitem
\bibitem[Yin \latin{et~al.}(2021)Yin, Yi, and Guo]{Yin2021High}
Yin,~Y.; Yi,~M.; Guo,~W. {High and anomalous thermal conductivity in monolayer
  $\text{MSi}_{\text{2}}\text{Z}_{\text{4}}$ semiconductors}. \emph{ACS Applied
  Materials \& Interfaces} \textbf{2021}, \emph{13}, 45907--45915\relax
\mciteBstWouldAddEndPuncttrue
\mciteSetBstMidEndSepPunct{\mcitedefaultmidpunct}
{\mcitedefaultendpunct}{\mcitedefaultseppunct}\relax
\EndOfBibitem
\bibitem[Yin \latin{et~al.}(2022)Yin, Gong, Yi, and Guo]{Yin2022Emerging}
Yin,~Y.; Gong,~Q.; Yi,~M.; Guo,~W. {Emerging versatile two-dimensional
  $\text{MoSi}_{\text{2}}\text{N}_{\text{4}}$ family}. \emph{Advanced
  Functional Materials} \textbf{2022}, 2214050\relax
\mciteBstWouldAddEndPuncttrue
\mciteSetBstMidEndSepPunct{\mcitedefaultmidpunct}
{\mcitedefaultendpunct}{\mcitedefaultseppunct}\relax
\EndOfBibitem
\bibitem[Li \latin{et~al.}(2019)Li, Ruan, and Zeng]{Li2019Intrinsic}
Li,~H.; Ruan,~S.; Zeng,~Y.~J. {Intrinsic van der Waals magnetic materials from
  bulk to the 2D limit: New frontiers of spintronics}. \emph{Advanced
  Materials} \textbf{2019}, \emph{31}, 1900065\relax
\mciteBstWouldAddEndPuncttrue
\mciteSetBstMidEndSepPunct{\mcitedefaultmidpunct}
{\mcitedefaultendpunct}{\mcitedefaultseppunct}\relax
\EndOfBibitem
\bibitem[Hou \latin{et~al.}(2019)Hou, Chen, Zhang, Wang, and Zhou]{Hou20192D}
Hou,~X.; Chen,~H.; Zhang,~Z.; Wang,~S.; Zhou,~P. {2D atomic crystals: A
  promising solution for next-generation data storage}. \emph{Advanced
  Electronic Materials} \textbf{2019}, \emph{5}, 1800944\relax
\mciteBstWouldAddEndPuncttrue
\mciteSetBstMidEndSepPunct{\mcitedefaultmidpunct}
{\mcitedefaultendpunct}{\mcitedefaultseppunct}\relax
\EndOfBibitem
\bibitem[Wang \latin{et~al.}(2020)Wang, Qi, and Qian]{Wang2020Electrically}
Wang,~H.; Qi,~J.; Qian,~X. {Electrically tunable high Curie temperature
  two-dimensional ferromagnetism in van der Waals layered crystals}.
  \emph{Applied Physics Letters} \textbf{2020}, \emph{117}, 083102\relax
\mciteBstWouldAddEndPuncttrue
\mciteSetBstMidEndSepPunct{\mcitedefaultmidpunct}
{\mcitedefaultendpunct}{\mcitedefaultseppunct}\relax
\EndOfBibitem
\bibitem[Xue \latin{et~al.}(2022)Xue, He, Gong, Yi, and Guo]{Xue2022Nonlinear}
Xue,~M.; He,~W.; Gong,~Q.; Yi,~M.; Guo,~W. {Nonlinear elasticity and
  strain-tunable magnetocalorics of antiferromagnetic monolayer
  $\text{MnPS}_{\text{3}}$}. \emph{Extreme Mechanics Letters} \textbf{2022},
  \emph{57}, 101900\relax
\mciteBstWouldAddEndPuncttrue
\mciteSetBstMidEndSepPunct{\mcitedefaultmidpunct}
{\mcitedefaultendpunct}{\mcitedefaultseppunct}\relax
\EndOfBibitem
\bibitem[Tang \latin{et~al.}(2023)Tang, Gong, and Yi]{Tang2023Spin}
Tang,~Z.~M.; Gong,~Q.; Yi,~M.-Y. Spin-selective contact type and strong Fermi
  level pinning at $\text{CrI}_{\text{3}}$/metal interface. \emph{Materials
  Today Nano} \textbf{2023}, \emph{22}, 100309\relax
\mciteBstWouldAddEndPuncttrue
\mciteSetBstMidEndSepPunct{\mcitedefaultmidpunct}
{\mcitedefaultendpunct}{\mcitedefaultseppunct}\relax
\EndOfBibitem
\bibitem[Li \latin{et~al.}(2019)Li, Jiang, Sivadas, Wang, Xu, Weber,
  Goldberger, Watanabe, Taniguchi, Fennie, {Fai Mak}, and Shan]{Li2019Pressure}
Li,~T.; Jiang,~S.; Sivadas,~N.; Wang,~Z.; Xu,~Y.; Weber,~D.; Goldberger,~J.~E.;
  Watanabe,~K.; Taniguchi,~T.; Fennie,~C.~J.; {Fai Mak},~K.; Shan,~J.
  {Pressure-controlled interlayer magnetism in atomically thin
  $\text{CrI}_{\text{3}}$}. \emph{Nature Materials} \textbf{2019}, \emph{18},
  1303--1308\relax
\mciteBstWouldAddEndPuncttrue
\mciteSetBstMidEndSepPunct{\mcitedefaultmidpunct}
{\mcitedefaultendpunct}{\mcitedefaultseppunct}\relax
\EndOfBibitem
\bibitem[Ubrig \latin{et~al.}(2019)Ubrig, Wang, Teyssier, Taniguchi, Watanabe,
  Giannini, Morpurgo, and Gibertini]{Ubrig2019Low}
Ubrig,~N.; Wang,~Z.; Teyssier,~J.; Taniguchi,~T.; Watanabe,~K.; Giannini,~E.;
  Morpurgo,~A.~F.; Gibertini,~M. {Low-temperature monoclinic layer stacking in
  atomically thin $\text{CrI}_{\text{3}}$ crystals}. \emph{2D Materials}
  \textbf{2019}, \emph{7}, 015007\relax
\mciteBstWouldAddEndPuncttrue
\mciteSetBstMidEndSepPunct{\mcitedefaultmidpunct}
{\mcitedefaultendpunct}{\mcitedefaultseppunct}\relax
\EndOfBibitem
\bibitem[Guo \latin{et~al.}(2021)Guo, Jin, Ye, Ye, Xie, Yang, Kim, Yan, Fu,
  Tian, Lei, Tsen, Sun, Yan, He, and Zhao]{Guo2021Structural}
Guo,~X. \latin{et~al.}  {Structural monoclinicity and its coupling to layered
  magnetism in few-layer $\text{CrI}_{\text{3}}$}. \emph{ACS Nano}
  \textbf{2021}, \emph{15}, 10444--10450\relax
\mciteBstWouldAddEndPuncttrue
\mciteSetBstMidEndSepPunct{\mcitedefaultmidpunct}
{\mcitedefaultendpunct}{\mcitedefaultseppunct}\relax
\EndOfBibitem
\bibitem[Zhang \latin{et~al.}(2018)Zhang, Zhao, Zhou, Xue, Ma, and
  Yang]{Zhang2018Strong}
Zhang,~J.; Zhao,~B.; Zhou,~T.; Xue,~Y.; Ma,~C.; Yang,~Z. {Strong magnetization
  and Chern insulators in compressed graphene/$\text{CrI}_{\text{3}}$ van der
  Waals heterostructures}. \emph{Physical Review B} \textbf{2018}, \emph{97},
  085401\relax
\mciteBstWouldAddEndPuncttrue
\mciteSetBstMidEndSepPunct{\mcitedefaultmidpunct}
{\mcitedefaultendpunct}{\mcitedefaultseppunct}\relax
\EndOfBibitem
\bibitem[Song \latin{et~al.}(2018)Song, Cai, Tu, Zhang, Huang, Wilson, Seyler,
  Zhu, Taniguchi, Watanabe, McGuire, Cobden, Xiao, Yao, and Xu]{Song2018Giant}
Song,~T.; Cai,~X.; Tu,~M. W.~Y.; Zhang,~X.; Huang,~B.; Wilson,~N.~P.;
  Seyler,~K.~L.; Zhu,~L.; Taniguchi,~T.; Watanabe,~K.; McGuire,~M.~A.;
  Cobden,~D.~H.; Xiao,~D.; Yao,~W.; Xu,~X. {Giant tunneling magnetoresistance
  in spin-filter van der Waals heterostructures}. \emph{Science} \textbf{2018},
  \emph{360}, 1214--1218\relax
\mciteBstWouldAddEndPuncttrue
\mciteSetBstMidEndSepPunct{\mcitedefaultmidpunct}
{\mcitedefaultendpunct}{\mcitedefaultseppunct}\relax
\EndOfBibitem
\bibitem[Gibertini \latin{et~al.}(2019)Gibertini, Koperski, Morpurgo, and
  Novoselov]{Gibertini2019Magnetic}
Gibertini,~M.; Koperski,~M.; Morpurgo,~A.~F.; Novoselov,~K.~S. {Magnetic 2D
  materials and heterostructures}. \emph{Nature Nanotechnology} \textbf{2019},
  \emph{14}, 408--419\relax
\mciteBstWouldAddEndPuncttrue
\mciteSetBstMidEndSepPunct{\mcitedefaultmidpunct}
{\mcitedefaultendpunct}{\mcitedefaultseppunct}\relax
\EndOfBibitem
\bibitem[Li \latin{et~al.}(2020)Li, Xu, Cheng, He, and Zhang]{Li2020Spin}
Li,~H.; Xu,~Y.~K.; Cheng,~Z.~P.; He,~B.~G.; Zhang,~W.~B. {Spin-dependent
  Schottky barriers and vacancy-induced spin-selective ohmic contacts in
  magnetic vdW heterostructures}. \emph{Physical Chemistry Chemical Physics}
  \textbf{2020}, \emph{22}, 9460--9466\relax
\mciteBstWouldAddEndPuncttrue
\mciteSetBstMidEndSepPunct{\mcitedefaultmidpunct}
{\mcitedefaultendpunct}{\mcitedefaultseppunct}\relax
\EndOfBibitem
\bibitem[Gong \latin{et~al.}(2020)Gong, Yi, and Xu]{Gong2020Electric}
Gong,~Q.; Yi,~M.; Xu,~B.~X. {Electric field induced magnetization reversal in
  magnet/insulator nanoheterostructure}. \emph{International Journal of Smart
  and Nano Materials} \textbf{2020}, \emph{11}, 298--309\relax
\mciteBstWouldAddEndPuncttrue
\mciteSetBstMidEndSepPunct{\mcitedefaultmidpunct}
{\mcitedefaultendpunct}{\mcitedefaultseppunct}\relax
\EndOfBibitem
\bibitem[Yao \latin{et~al.}(2021)Yao, Zhan, Sendeku, Yu, Dajan, Li, Wang, Zhu,
  Wang, Wang, and He]{Yao2021Recent}
Yao,~Y.; Zhan,~X.; Sendeku,~M.~G.; Yu,~P.; Dajan,~F.~T.; Li,~N.; Wang,~J.;
  Zhu,~C.; Wang,~F.; Wang,~Z.; He,~J. {Recent progress on emergent
  two-dimensional magnets and heterostructures}. \emph{Nanotechnology}
  \textbf{2021}, \emph{32}, 472001\relax
\mciteBstWouldAddEndPuncttrue
\mciteSetBstMidEndSepPunct{\mcitedefaultmidpunct}
{\mcitedefaultendpunct}{\mcitedefaultseppunct}\relax
\EndOfBibitem
\bibitem[Tan \latin{et~al.}(2021)Tan, Ding, Du, and Fu]{Tan2021Spin}
Tan,~X.; Ding,~L.; Du,~G.~F.; Fu,~H.~H. {Spin caloritronics in two-dimensional
  $\text{CrI}_{\text{3}}$/$\text{NiCl}_{\text{2}}$ van der Waals
  heterostructures}. \emph{Physical Review B} \textbf{2021}, \emph{103},
  115415\relax
\mciteBstWouldAddEndPuncttrue
\mciteSetBstMidEndSepPunct{\mcitedefaultmidpunct}
{\mcitedefaultendpunct}{\mcitedefaultseppunct}\relax
\EndOfBibitem
\bibitem[Petrov \latin{et~al.}(2019)Petrov, Silkin, Menshchikova, and
  Chulkov]{Petrov2019Cr}
Petrov,~E.~K.; Silkin,~I.~V.; Menshchikova,~T.~V.; Chulkov,~E.~V.
  {Cr-containing ferromagnetic film–topological insulator heterostructures as
  promising materials for the quantum anomalous hall effect}. \emph{JETP
  Letters} \textbf{2019}, \emph{109}, 121--125\relax
\mciteBstWouldAddEndPuncttrue
\mciteSetBstMidEndSepPunct{\mcitedefaultmidpunct}
{\mcitedefaultendpunct}{\mcitedefaultseppunct}\relax
\EndOfBibitem
\bibitem[Gao \latin{et~al.}(2022)Gao, Li, and Zhu]{Gao2022Prediction}
Gao,~Y.; Li,~H.; Zhu,~W. {Prediction of quantum anomalous Hall effect in
  $\text{CrI}_{\text{3}}$/$\text{ScCl}_{\text{2}}$ bilayer heterostructure}.
  \emph{Chinese Physics B} \textbf{2022}, \emph{31}, 107304\relax
\mciteBstWouldAddEndPuncttrue
\mciteSetBstMidEndSepPunct{\mcitedefaultmidpunct}
{\mcitedefaultendpunct}{\mcitedefaultseppunct}\relax
\EndOfBibitem
\bibitem[Zhao \latin{et~al.}(2019)Zhao, Zhang, Yuan, and
  Chen]{Zhao2019Nonvolatile}
Zhao,~Y.; Zhang,~J.~J.; Yuan,~S.; Chen,~Z. {Nonvolatile electrical control and
  heterointerface-induced half-metallicity of 2D ferromagnets}. \emph{Advanced
  Functional Materials} \textbf{2019}, \emph{29}, 1901420\relax
\mciteBstWouldAddEndPuncttrue
\mciteSetBstMidEndSepPunct{\mcitedefaultmidpunct}
{\mcitedefaultendpunct}{\mcitedefaultseppunct}\relax
\EndOfBibitem
\bibitem[Chakraborty and Ravikumar(2021)Chakraborty, and
  Ravikumar]{Chakraborty2021Substrate}
Chakraborty,~S.; Ravikumar,~A. {Substrate induced electronic phase transitions
  of $\text{CrI}_{\text{3}}$ based van der Waals heterostructures}.
  \emph{Scientific Reports} \textbf{2021}, \emph{11}, 198\relax
\mciteBstWouldAddEndPuncttrue
\mciteSetBstMidEndSepPunct{\mcitedefaultmidpunct}
{\mcitedefaultendpunct}{\mcitedefaultseppunct}\relax
\EndOfBibitem
\bibitem[Wang \latin{et~al.}(2021)Wang, Qin, Wang, Teketel, Yu, Luo, Xu, and
  Lin]{Wang2021Heterointerface}
Wang,~G.; Qin,~W.; Wang,~S.; Teketel,~B.~S.; Yu,~W.; Luo,~T.; Xu,~B.; Lin,~B.
  {$\text{CrI}_{\text{3}}$/$\text{Y}_\text{2}\text{CH}_\text{2}$:Heterointerface-induced
  stable half-metallicity of two-dimensional $\text{CrI}_{\text{3}}$ monolayer
  ferromagnets}. \emph{ACS Applied Materials and Interfaces} \textbf{2021},
  \emph{13}, 16694--16703\relax
\mciteBstWouldAddEndPuncttrue
\mciteSetBstMidEndSepPunct{\mcitedefaultmidpunct}
{\mcitedefaultendpunct}{\mcitedefaultseppunct}\relax
\EndOfBibitem
\bibitem[Chen \latin{et~al.}(2019)Chen, Huang, Sun, Ding, Jena, and
  Kan]{Chen2019Boosting}
Chen,~S.; Huang,~C.; Sun,~H.; Ding,~J.; Jena,~P.; Kan,~E. {Boosting the Curie
  temperature of two-dimensional semiconducting $\text{CrI}_{\text{3}}$
  monolayer through van der Waals heterostructures}. \emph{The Journal of
  Physical Chemistry C} \textbf{2019}, \emph{123}, 17987--17993\relax
\mciteBstWouldAddEndPuncttrue
\mciteSetBstMidEndSepPunct{\mcitedefaultmidpunct}
{\mcitedefaultendpunct}{\mcitedefaultseppunct}\relax
\EndOfBibitem
\bibitem[Hu \latin{et~al.}(2021)Hu, Tan, Wu, Zhang, and Fan]{Hu2021Exploring}
Hu,~J.~K.; Tan,~J.~X.; Wu,~D.; Zhang,~Z.~H.; Fan,~Z.~Q. {Exploring magnetic
  stability and valley splitting on $\text{CrI}_{\text{3}}$/SiC van der Waals
  heterostructure}. \emph{Applied Surface Science} \textbf{2021}, \emph{560},
  149858\relax
\mciteBstWouldAddEndPuncttrue
\mciteSetBstMidEndSepPunct{\mcitedefaultmidpunct}
{\mcitedefaultendpunct}{\mcitedefaultseppunct}\relax
\EndOfBibitem
\bibitem[Yu \latin{et~al.}(2022)Yu, Luo, Zhang, Wu, Jia, Yang, Cai, Song,
  Zhang, and Zhang]{Yu2022Strain}
Yu,~W.; Luo,~W.; Zhang,~X.; Wu,~Y.; Jia,~X.; Yang,~X.; Cai,~X.; Song,~A.;
  Zhang,~Z.; Zhang,~W.~B. {Strain and electric field dependent spin
  polarization in two-dimensional arsenene/$\text{CrI}_{\text{3}}$
  heterostructure}. \emph{Journal of Alloys and Compounds} \textbf{2022},
  \emph{912}, 165093\relax
\mciteBstWouldAddEndPuncttrue
\mciteSetBstMidEndSepPunct{\mcitedefaultmidpunct}
{\mcitedefaultendpunct}{\mcitedefaultseppunct}\relax
\EndOfBibitem
\bibitem[Yang \latin{et~al.}(2021)Yang, Hu, Shen, Krasheninnikov, Chen, and
  Sun]{Yang2021Enhancing}
Yang,~Q.; Hu,~X.; Shen,~X.; Krasheninnikov,~A.~V.; Chen,~Z.; Sun,~L. {Enhancing
  ferromagnetism and tuning electronic properties of $\text{CrI}_{\text{3}}$
  monolayers by adsorption of transition-metal atoms}. \emph{ACS Applied
  Materials \& Interfaces} \textbf{2021}, \emph{13}, 21593--21601\relax
\mciteBstWouldAddEndPuncttrue
\mciteSetBstMidEndSepPunct{\mcitedefaultmidpunct}
{\mcitedefaultendpunct}{\mcitedefaultseppunct}\relax
\EndOfBibitem
\bibitem[Jiang \latin{et~al.}(2018)Jiang, Li, Liao, Zhao, and
  Zhong]{Jiang2018Spin}
Jiang,~P.; Li,~L.; Liao,~Z.; Zhao,~Y.~X.; Zhong,~Z. {Spin direction-controlled
  electronic band structure in two-dimensional ferromagnetic
  $\text{CrI}_{\text{3}}$}. \emph{Nano Letters} \textbf{2018}, \emph{18},
  3844--3849\relax
\mciteBstWouldAddEndPuncttrue
\mciteSetBstMidEndSepPunct{\mcitedefaultmidpunct}
{\mcitedefaultendpunct}{\mcitedefaultseppunct}\relax
\EndOfBibitem
\bibitem[Liu \latin{et~al.}(2018)Liu, Shi, Lu, and Anantram]{Liu2018Analysis}
Liu,~J.; Shi,~M.; Lu,~J.; Anantram,~M.~P. {Analysis of
  electrical-field-dependent Dzyaloshinskii-Moriya interaction and
  magnetocrystalline anisotropy in a two-dimensional ferromagnetic monolayer}.
  \emph{Physical Review B} \textbf{2018}, \emph{97}, 8--10\relax
\mciteBstWouldAddEndPuncttrue
\mciteSetBstMidEndSepPunct{\mcitedefaultmidpunct}
{\mcitedefaultendpunct}{\mcitedefaultseppunct}\relax
\EndOfBibitem
\bibitem[He \latin{et~al.}(2023)He, Yin, Gong, Evans, Gutfleisch, Xu, Yi, and
  Guo]{He2023Giant}
He,~W.; Yin,~Y.; Gong,~Q.; Evans,~R.~F.; Gutfleisch,~O.; Xu,~B.; Yi,~M.;
  Guo,~W. Giant magnetocaloric effect in magnets down to the monolayer limit.
  \emph{arXiv preprint} \textbf{2023}, \relax
\mciteBstWouldAddEndPunctfalse
\mciteSetBstMidEndSepPunct{\mcitedefaultmidpunct}
{}{\mcitedefaultseppunct}\relax
\EndOfBibitem
\bibitem[Liu and Petrovic(2018)Liu, and Petrovic]{Liu2018Anisotropic}
Liu,~Y.; Petrovic,~C. {Anisotropic magnetocaloric effect in single crystals of
  $\text{CrI}_{\text{3}}$}. \emph{Physical Review B} \textbf{2018}, \emph{97},
  174418\relax
\mciteBstWouldAddEndPuncttrue
\mciteSetBstMidEndSepPunct{\mcitedefaultmidpunct}
{\mcitedefaultendpunct}{\mcitedefaultseppunct}\relax
\EndOfBibitem
\bibitem[Tran \latin{et~al.}(2022)Tran, Momida, ichiro Matsushita, Shirai, and
  Oguchi]{Tran2022Insight}
Tran,~H.~B.; Momida,~H.; ichiro Matsushita,~Y.; Shirai,~K.; Oguchi,~T. {Insight
  into anisotropic magnetocaloric effect of $\text{CrI}_{\text{3}}$}.
  \emph{Acta Materialia} \textbf{2022}, \emph{231}, 117851\relax
\mciteBstWouldAddEndPuncttrue
\mciteSetBstMidEndSepPunct{\mcitedefaultmidpunct}
{\mcitedefaultendpunct}{\mcitedefaultseppunct}\relax
\EndOfBibitem
\bibitem[Kresse and Furthm\"uller(1996)Kresse, and
  Furthm\"uller]{Kresse1996Efficient}
Kresse,~G.; Furthm\"uller,~J. {Efficient iterative schemes for $ab~initio$
  total-energy calculations using a plane-wave basis set}. \emph{Physical
  Review B} \textbf{1996}, \emph{54}, 11169--11186\relax
\mciteBstWouldAddEndPuncttrue
\mciteSetBstMidEndSepPunct{\mcitedefaultmidpunct}
{\mcitedefaultendpunct}{\mcitedefaultseppunct}\relax
\EndOfBibitem
\bibitem[Kresse and Joubert(1999)Kresse, and Joubert]{Kresse1999From}
Kresse,~G.; Joubert,~D. {From ultrasoft pseudopotentials to the projector
  augmented-wave method}. \emph{Physical Review B} \textbf{1999}, \emph{59},
  1758--1775\relax
\mciteBstWouldAddEndPuncttrue
\mciteSetBstMidEndSepPunct{\mcitedefaultmidpunct}
{\mcitedefaultendpunct}{\mcitedefaultseppunct}\relax
\EndOfBibitem
\bibitem[Perdew \latin{et~al.}(1996)Perdew, Burke, and
  Ernzerhof]{Perdew1996Generalized}
Perdew,~J.~P.; Burke,~K.; Ernzerhof,~M. {Generalized Gradient Approximation
  Made Simple}. \emph{Physical Review Letters} \textbf{1996}, \emph{77},
  3865--3868\relax
\mciteBstWouldAddEndPuncttrue
\mciteSetBstMidEndSepPunct{\mcitedefaultmidpunct}
{\mcitedefaultendpunct}{\mcitedefaultseppunct}\relax
\EndOfBibitem
\bibitem[Neugebauer and Scheffler(1992)Neugebauer, and
  Scheffler]{Neugebauer1992Adsorbate}
Neugebauer,~J.; Scheffler,~M. {Adsorbate-substrate and adsorbate-adsorbate
  interactions of Na and K adlayers on Al(111)}. \emph{Physical Review B}
  \textbf{1992}, \emph{46}, 16067--16080\relax
\mciteBstWouldAddEndPuncttrue
\mciteSetBstMidEndSepPunct{\mcitedefaultmidpunct}
{\mcitedefaultendpunct}{\mcitedefaultseppunct}\relax
\EndOfBibitem
\bibitem[Monkhorst and Pack(1976)Monkhorst, and Pack]{Monkhorst1976Special}
Monkhorst,~H.~J.; Pack,~J.~D. {Special points for Brillouin-zone integrations}.
  \emph{Physical Review B} \textbf{1976}, \emph{13}, 5188--5192\relax
\mciteBstWouldAddEndPuncttrue
\mciteSetBstMidEndSepPunct{\mcitedefaultmidpunct}
{\mcitedefaultendpunct}{\mcitedefaultseppunct}\relax
\EndOfBibitem
\bibitem[Vaz \latin{et~al.}(2008)Vaz, Bland, and Lauhoff]{Vaz2008Magnetism}
Vaz,~C. A.~F.; Bland,~J. A.~C.; Lauhoff,~G. {Magnetism in ultrathin film
  structures}. \emph{Reports on Progress in Physics} \textbf{2008}, \emph{71},
  056501\relax
\mciteBstWouldAddEndPuncttrue
\mciteSetBstMidEndSepPunct{\mcitedefaultmidpunct}
{\mcitedefaultendpunct}{\mcitedefaultseppunct}\relax
\EndOfBibitem
\bibitem[Skubic \latin{et~al.}(2008)Skubic, Hellsvik, Nordström, and
  Eriksson]{Skubic2008A}
Skubic,~B.; Hellsvik,~J.; Nordström,~L.; Eriksson,~O. {A method for atomistic
  spin dynamics simulations: Implementation and examples}. \emph{Journal of
  Physics: Condensed Matter} \textbf{2008}, \emph{20}, 315203\relax
\mciteBstWouldAddEndPuncttrue
\mciteSetBstMidEndSepPunct{\mcitedefaultmidpunct}
{\mcitedefaultendpunct}{\mcitedefaultseppunct}\relax
\EndOfBibitem
\bibitem[Evans \latin{et~al.}(2014)Evans, Fan, Chureemart, Ostler, Ellis, and
  Chantrell]{Evans2014Atomistic}
Evans,~R.~F.; Fan,~W.~J.; Chureemart,~P.; Ostler,~T.~A.; Ellis,~M.~O.;
  Chantrell,~R.~W. {Atomistic spin model simulations of magnetic
  nanomaterials}. \emph{Journal of Physics: Condensed Matter} \textbf{2014},
  \emph{26}, 103202\relax
\mciteBstWouldAddEndPuncttrue
\mciteSetBstMidEndSepPunct{\mcitedefaultmidpunct}
{\mcitedefaultendpunct}{\mcitedefaultseppunct}\relax
\EndOfBibitem
\bibitem[Gong \latin{et~al.}(2019)Gong, Yi, Evans, Xu, and
  Gutfleisch]{Gong2019Calculating}
Gong,~Q.; Yi,~M.; Evans,~R.~F.; Xu,~B.~X.; Gutfleisch,~O. {Calculating
  temperature-dependent properties of
  $\text{Nd}_{\text{2}}\text{Fe}_{\text{14}}\text{B}$ permanent magnets by
  atomistic spin model simulations}. \emph{Physical Review B} \textbf{2019},
  \emph{99}, 214409\relax
\mciteBstWouldAddEndPuncttrue
\mciteSetBstMidEndSepPunct{\mcitedefaultmidpunct}
{\mcitedefaultendpunct}{\mcitedefaultseppunct}\relax
\EndOfBibitem
\bibitem[Gong \latin{et~al.}(2019)Gong, Yi, and Xu]{Gong2019Multiscale}
Gong,~Q.; Yi,~M.; Xu,~B.~X. {Multiscale simulations toward calculating
  coercivity of Nd-Fe-B permanent magnets at high temperatures}. \emph{Physical
  Review Materials} \textbf{2019}, \emph{3}, 84406\relax
\mciteBstWouldAddEndPuncttrue
\mciteSetBstMidEndSepPunct{\mcitedefaultmidpunct}
{\mcitedefaultendpunct}{\mcitedefaultseppunct}\relax
\EndOfBibitem
\bibitem[Pecharsky and Gschneidner(1999)Pecharsky, and
  Gschneidner]{Pecharsky1999Magnetocaloric}
Pecharsky,~V.~K.; Gschneidner,~K.~A. {Magnetocaloric effect from indirect
  measurements: Magnetization and heat capacity}. \emph{Journal of Applied
  Physics} \textbf{1999}, \emph{86}, 565--575\relax
\mciteBstWouldAddEndPuncttrue
\mciteSetBstMidEndSepPunct{\mcitedefaultmidpunct}
{\mcitedefaultendpunct}{\mcitedefaultseppunct}\relax
\EndOfBibitem
\bibitem[Gschneidner and Pecharsky(2000)Gschneidner, and
  Pecharsky]{Gschneidner2000Magnetocaloric}
Gschneidner,~K.~A.; Pecharsky,~V.~K. {Magnetocaloric materials}. \emph{Annual
  Review of Materials Science} \textbf{2000}, \emph{30}, 387--429\relax
\mciteBstWouldAddEndPuncttrue
\mciteSetBstMidEndSepPunct{\mcitedefaultmidpunct}
{\mcitedefaultendpunct}{\mcitedefaultseppunct}\relax
\EndOfBibitem
\end{mcitethebibliography}
\end{spacing}
\includepdfmerge{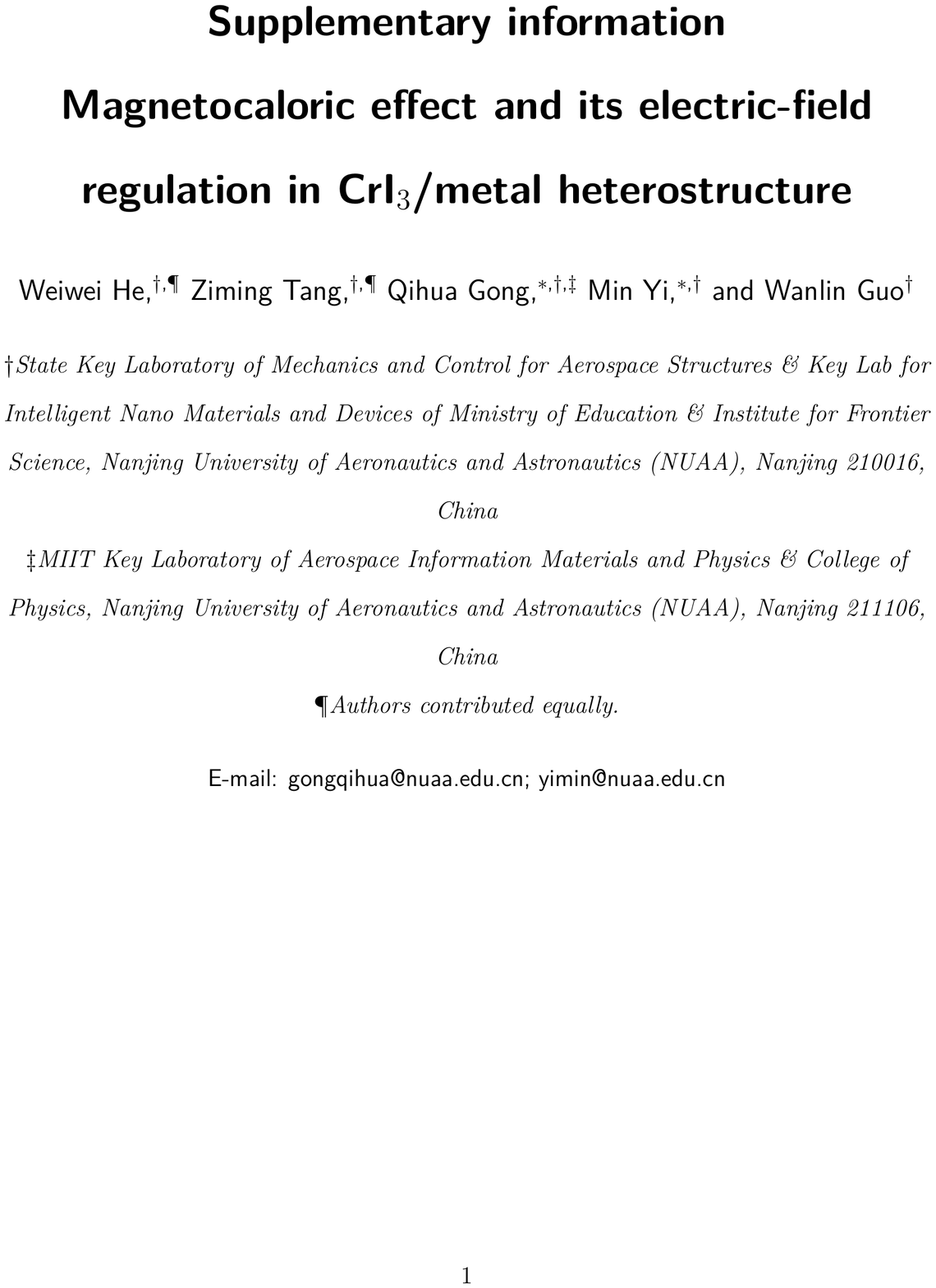,1-9}
\end{document}